\documentclass[aps,twocolumn,showpacs,amsmath,amssymb]{revtex4}

\usepackage{graphicx} 
\usepackage[dvips]{epsfig}

\newcommand{\be}{\begin{eqnarray}}
\newcommand{\ee}{\end{eqnarray}}

\newcommand{\Eb}{\protect{{\bf E}}}

\newcommand{\Gb}{\protect{{\bf G}}}

\newcommand{\rb}{\protect{{\bf r}}}

\newcommand{\kb}{\protect{{\bf k}}}
\newcommand{\Kb}{\protect{{\bf K}}}
\newcommand{\db}{\protect{{\bf d}}}

\renewcommand{\Im}{\text{Im}}

\begin{document} 

\title{Sub- and super-radiance over macroscopic distances using 
a perfect lens with negative refraction}
\author{J\"urgen K\"astel and Michael Fleischhauer}
\affiliation{Fachbereich Physik, Technische Universit\"{a}t 
Kaiserslautern, D-67663 Kaiserslautern, 
Germany} 
\date{\today} 

\begin{abstract} 
Two atoms put at the foci of a perfect lens
[J.B. Pendry, Phys. Ref. Lett. {\bf 85}, 3966 (2000)] are shown to 
exhibit perfect sub- and super-radiance even over macroscopic 
distances limited only by the propagation length in the free-space decay time.
If the left-handed material forming
the perfect lens has nearly constant negative refraction and
vanishing absorption over a spectral range larger than the natural
linewidth, the imaginary part of the retarded Greens-function between the
two focal points is identical to the one at the same spatial position and 
the atoms undergo a Markovian dynamics.
Collective decay rates and level shifts are calculated from the Greens-function
of the Veselago-Pendry lens and limitation as well as
potential applications are discussed.
\end{abstract} 
 
 
\maketitle

%%%%%%%%%%%%%%%%%%%%%%%%%%%%%%%%%%%%%%%%%%%%%%%%%%%%%%%%%%%%%%%%%%%%%%%%%%
%%%%%%%%%%%%%%%%%%%%%%%%%%%%%%%%%%%%%%%%%%%%%%%%%%%%%%%%%%%%%%%%%%%%%%%%%%

Negative refraction of electromagnetic radiation 
in materials with simultaneous 
negative dielectric positivity $\varepsilon$ and 
magnetic permeability $\mu$ was
first predicted by Veselago \cite{Veselago68}. 
These so-called left-handed media attracted 
much attention recently because of possible realizations in 
metamaterials \cite{Smith00a,Shelby01sci,Marques02,Grbic02,Parazzoli03} and 
their application for a lens without 
diffraction limitations \cite{Pendry00}. An infinite parallel slab of lossless
left-handed material of thickness $d$ collects all
plane waves from a point source on one side of the slab not too far 
away from the surface in a focal point
on the other side. If the medium has a refractive index of $n=-1$ the
distance between the two foci is $2d$. As pointed out by Pendry 
\cite{Pendry00}, the lens formed by the slab 
is perfect in the sense that the amplitudes
of evanescent waves emerging from the source are amplified in the left
handed medium (LHM) and exactly reproduced at the focal point thus leading to 
an image not limited by diffraction. This raises the question what 
happens to two atoms with radiative transitions inside the frequency 
range of negative refraction put in the focal points of a 
Veselago-Pendry lens. 

We here show that the
imaginary part of the retarded Greens-function between the two focal points
is under ideal conditions identical to the free-space Greens-function at
the same positions. As a consequence there occurs 
perfect sub- and super-radiance
\cite{Dicke54} of the two atoms as well as dipole-dipole shifts
even over distances large compared 
to the transition 
wavelength.
The strong radiative coupling 
persists as long as the distance between the focal points is
smaller than the propagation length during the free-space radiative 
decay time. For larger distances retardation and memory effects become 
important and the two-atom system can no longer be described by a
master equation. 

\begin{figure}[ht] 
  \begin{center} 
    \includegraphics[width=6cm]{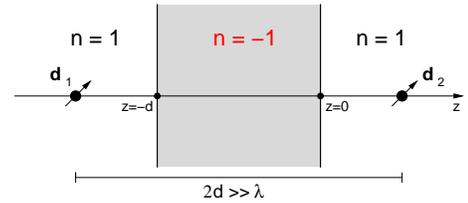} 
    \caption{Two atoms put into the focal points of a
Veselago-Pendry lens with $n=-1$. Focal points are all pairs of positions
at the two sides of the slab with distance $2d$.
The spatial regions $z>0$ (vacuum), $-d\le z\le 0$ (LHM), and $z<-d$ (vacuum)
are denoted by the numbers $0,1,2$ respectively.}
    \label{fig1} 
  \end{center} 
\end{figure}

Let us consider an infinitely extended slab of homogeneous 
LHM of thickness $d$ and two atoms  put in 
two focal points of the 
Veselago-Pendry lens formed by the slab as shown in fig.\ref{fig1}.
The distance between the focal points is $d(1-n)$, $n<0$ being the
refractive index of the LHM.
The atoms are two-level systems with
ground states $|1\rangle$ and excited states $|2\rangle$ and common
transition frequency $\omega_0$.
The dipole vectors of the atoms are denoted by $\mathbf{d_1}$ 
and $\mathbf{d_2}$.

%%%%%%%%%%%%%%%%%%%%%%%%%%%%%%%%%%%%%%%%%%%%%%%%%%%%%%%%%%%%%%%%%%%%%%%%%%

%%%%%%%%%%%%%%%%%%%%%%%%%%%%%%%%%%%%%%%%%%%%%%%%%%%%%%%%%%%%%%%%%%%%%%%%%%

The coupling of the atoms to the quantized radiation field is described 
by the interaction Hamiltonian in
dipole approximation 
\begin{equation}
H_{\text{WW}} = -\hat\db_1\cdot\hat\Eb(\rb_1) -\hat\db_1\cdot\hat\Eb(\rb_2),
\end{equation}
where $\hat\Eb(\rb)$ is the operator of the electric field in the
presence of the LHM. Employing the usual Born-Markov and rotating-wave 
approximations 
one can derive a master equation for the two-atom density matrix in the
interaction picture
\begin{equation}
\label{Liouville}
\begin{split}
\dot\rho = & 
-\sum_{k,l=1}^2\frac{\Gamma(\rb_k,\rb_l)}{2}\Bigl(\hat\sigma^{\dagger}_l
\hat\sigma_k\rho+\rho\hat\sigma^{\dagger}_l
\hat\sigma_k-2\hat\sigma_k\rho\hat\sigma^{\dagger}_l\Bigr) \\
& +i\sum_{k,l=1}^2\delta\omega(\rb_k,\rb_l)
\left[\hat\sigma^{\dagger}_l\hat\sigma_k,\rho\right].
\end{split}
\end{equation}
Here $\hat\sigma_k
=|1\rangle_{kk}\langle 2|$ are the flip operators of the $k$th atom. The rates
 $\Gamma (\rb_k,\rb_l)$ describe the radiative decay of the two two-level 
atoms. $\Gamma(\rb_k,\rb_k)$ corresponds to the single-particle decay rates
of an atom at position $\rb_k$ and $\Gamma(\rb_1,\rb_2)$ 
describes the dissipative cross coupling. Both quantities are determined by
the imaginary part of the retarded Greens-tensor of the electric field 
at the atomic transition
frequency $\Gb(\rb_k,\rb_l,\omega_0)=G_{\mu\nu}(\rb_k,\rb_l,\omega_0)
\hat{\boldmath{\mu}}\circ \hat{\boldmath{\nu}} $, ``$\circ$'' denoting 
a tensorial 
product and $\hat{\mathbf{\mu}}$ and $\hat{\mathbf{\nu}}$ are unit vectors
\cite{Dung03}.
\begin{equation}
\Gamma(\rb_k,\rb_l)= \frac{2\omega^2d_\mu d_\nu}
{\hbar\varepsilon_0c^2}\Im \left[G_{\mu\nu} (\rb_k,\rb_l,\omega_0)\right].
\label{decay-rate}
\end{equation}
$\delta\omega(\rb_k,\rb_k)$ is the single-atom Lamb shift and
$\delta\omega(\rb_1,\rb_2)$ describes the
radiative dipole-dipole shift. It is well known that the Lamb shift is not
correctly described by the two-level model. Furthermore its explicit expression
as given below diverges and a renormalization is needed. 
The Lamb shift is however of no relevance for the present discussion and will
be ignored, i.e. it is assumed to be included in the
bare transition frequency. In contrast the dipole-dipole shift 
can correctly be calculated within the present model, when the free-field
part is subtracted. This is because the Veselago-Pendry lens can only lead to
contributions within a finite frequency window \cite{Veselago68}.
Subtracting the free-field contribution
one finds
\begin{equation}
\delta\omega= \frac{d_\mu d_\nu}{\hbar\pi\varepsilon_0}
{\mathcal P}\int_0^\infty d\omega\frac{\omega^2}{c^2} 
\frac{\Im \left[\Delta G_{\mu\nu}(\rb_1,\rb_2,\omega)\right]}{\omega-\omega_0},
\label{shift}
\end{equation}
where $\Delta G_{\mu\nu}(\rb_1,\rb_2,\omega) = G_{\mu\nu}(\rb_1,\rb_2,\omega)
-G_{\mu\nu}^0(\rb_1,\rb_2,\omega)$,
$G_{\mu\nu}^0(\rb_1,\rb_2,\omega)$ being the components of the free-field
retarded Greens-tensor.

The master equation \eqref{Liouville} for the two-atom system can be
written in a diagonal form introducing a basis of symmetric and
antisymmetric states $|11\rangle,|22\rangle$ and $|{\rm s}\rangle
=\bigl(|12\rangle +|21\rangle\bigr)/\sqrt{2}$
and $|{\rm a}\rangle
=\bigl(|12\rangle -|21\rangle\bigr)/\sqrt{2}$. This yields for the populations
\begin{eqnarray}
\dot\rho_{22} &=& -2 \Gamma_{11} \rho_{22},\\
\dot\rho_{\rm ss} &=& +(\Gamma_{11}+\Gamma_{12}) \rho_{22}
-(\Gamma_{11}+\Gamma_{12}) \rho_{\rm ss},\label{symm}\\
\dot\rho_{\rm {\rm aa}} &=& +(\Gamma_{11}-\Gamma_{12}) \rho_{22}
-(\Gamma_{11}-\Gamma_{12}) \rho_{{\rm aa}},\label{asymm}\\
\dot\rho_{11} &=& + (\Gamma_{11}+\Gamma_{12})\rho_{\rm ss}
+(\Gamma_{11}-\Gamma_{12})\rho_{{\rm aa}},
\end{eqnarray}
where $\Gamma_{11} =\Gamma(\rb,\rb)$ and $\Gamma_{12}=\Gamma(\rb_1,\rb_2)$.
In addition there is a level shift of the symmetric and antisymmetric
states $|{\rm s}\rangle$ and $|{\rm a}\rangle$ below or above the 
single atom energy by the dipole-dipole shift $\delta\omega$, given in 
eq.\eqref{shift}.

The retarded Greens-function corresponding to a slab  
with a homogeneous and linear magneto-dielectric medium (fig.\ref{fig1}) 
can be calculated
by a plane wave decomposition. Following \cite{Tsang85} one finds for the two
positions $\rb$ and $\rb^\prime$ in vacuum on the same side of the lens
\begin{eqnarray}
&&\Gb^{00}(\rb,\rb',\omega) =  
\frac{i}{8\pi^2}\int d^2 k_\perp \frac{1}{k_z} \Bigl[
\label{Green00}\\
&& \qquad 
 \bigl( R^{\rm TE}\hat {\bf e}(k_z) e^{i\kb\cdot\rb} 
+ \hat {\bf e}(-k_z) e^{i\Kb\cdot\rb} \bigr) 
\circ \hat {\bf e}(-k_z) e^{-i\Kb\cdot\rb'} \nonumber\\
&&\quad + \bigl(R^{\rm TM} \hat {\bf h}(k_z) e^{i\kb\cdot\rb} 
+ \hat {\bf h}(-k_z) e^{i\Kb\cdot\rb} \bigr) \circ
\hat {\bf h}(-k_z) e^{-i\Kb\rb'} \Bigr],\nonumber
\end{eqnarray}
where $z\le z^\prime$ has been assumed.
For $\rb$ and $\rb^\prime$ being in vacuum on different sides 
of the lens one finds
\begin{eqnarray}
&&\Gb^{20}(\rb,\rb',\omega)=  
\frac{i}{8\pi^2}\int d^2 k_\perp \frac{1}{k_z} \Bigl[
\nonumber\\
&&\qquad  T^{\rm TE}\hat {\bf e}(-k_z) e^{i\Kb\cdot\rb} 
\circ \hat {\bf e}(-k_z) e^{-i\Kb\cdot\rb'} \label{Green20}\\
&&\quad + T^{\rm TM} \hat {\bf h}(-k_z) e^{i\Kb\cdot\rb} \circ
\hat {\bf h}(-k_z) e^{-i\Kb\rb'} \Bigr].\nonumber 
\end{eqnarray}
The superscripts $0,1,2$ at the Greens-functions 
denote the zones of positions $\rb$ and $\rb'$: 
$z>0$ (vacuum), $-d\le z\le 0$ (LHM), and $z<-d$ (vacuum)
respectively.
We here have used the definitions $k^2 ={\omega^2}/{c^2}$, 
$k_z =\sqrt{(k^2-k_\perp^2)}$ and
$d^2k_\perp = dk_xdk_y$. Furthermore
$\Kb \equiv k_x\hat {\bf x}+k_y\hat {\bf y}-k_z\hat {\bf z}$ and we have
introduced the orthogonal unit vectors 
$\hat {\bf e} = {\kb\times\hat {\bf z}}/{\left|\kb\times\hat {\bf z}\right|}$
and $\hat {\bf h} = {p}\hat {\bf e}\times \kb/\left|k\right|$, where $p=1$ for
a normal medium and $p=-1$ for a LHM.
$R^{\rm TE}, R^{\rm TM}$ and $T^{\rm TE}, T^{\rm TM}$ are the reflection and
transmission functions of the lens for transverse electric and transverse
magnetic modes. They read
\begin{eqnarray}
\label{RTE}
R^{\rm TE} &=& \frac{R_{01}+R_{12}e^{i2k_{1z}d}}{1+R_{01}R_{12}e^{i2k_{1z}d}},\\
\label{RTM}
R^{\rm TM} &=& \frac{S_{01}+S_{12}e^{i2k_{1z}d}}{1+S_{01}S_{12}e^{i2k_{1z}d}}
\end{eqnarray}
and correspondingly
\begin{equation}
\label{TTE}
T^{\rm TE} = \frac{2\mu k_z}{\mu k_z+k_{1z}}\frac{1+R_{12}}
{1+R_{01}R_{12}e^{i2k_{1z}d}}e^{i(k_{1z}-k_z)d},
\end{equation}
\begin{equation}
\label{TTM}
T^{\rm TM} = \frac{2\varepsilon k_z}{\varepsilon k_z+k_{1z}}
\frac{1+S_{12}}{1+S_{01}S_{12}e^{i2k_{1z}d}}e^{i(k_{1z}-k_z)d}.
\end{equation}
Here $k_{1z}=\sqrt{k_1^2-k_\perp^2}$ and 
$k_1^2 = \varepsilon(\omega)\mu(\omega)  \omega^2/c^2$.
$R_{ij}$ and $S_{ij}$ are the reflection coefficients at the boundaries bet\-ween
media $i$ and $j$ for TE and TM modes respectively.
\begin{equation}
\label{RS}
R_{ij} = \frac{\mu_jk_{iz}-\mu_ik_{jz}}{\mu_jk_{iz}+\mu_ik_{jz}}
\qquad S_{ij} = \frac{\varepsilon_jk_{iz}-\varepsilon_ik_{jz}}
{\varepsilon_jk_{iz}+\varepsilon_ik_{jz}}
\end{equation}
The indexes $i,j\in\{0,1,2\}$ denote the region outside or inside the lens,
i.e. $k_0^2=k_2^2=k^2\equiv\omega^2/c^2$ and 
$k_1^2=\varepsilon(\omega)\mu(\omega)  \omega^2/c^2$.

From expressions \eqref{Green00} and \eqref{Green20} one can calculate 
Im$[\Gb(\rb_k,\rb_l,\omega_0)]$ for an ideal Veselago-Pendry-lens, i.e. for
infinite transversal extension and a lossless medium with $n(\omega_0)=-1$.
Since in this case $R^{\rm TE}=R^{\rm TM}=0$ one finds e.g.
Im$[\Gb^{00}(\rb,\rb,\omega_0)]= (k/6\pi)\, {\boldmath{\hat 1}}$, 
i.e. exactly the free-space
value. Most importantly one finds that for all points $\rb'$ in region 2 
$(z^\prime\le -d)$
\begin{eqnarray}
\textrm{Im}\Bigl[\Gb^{20}(\rb^\prime,\rb,\omega_0)\Bigr]=
\textrm{Im}\Bigl[\Gb^{00}(\rb^\prime+2 d{\mathbf{\hat z}},\rb,
\omega_0)\Bigr],
\end{eqnarray}
since 
$T^{\rm TE}=T^{\rm TM}=e^{ik_{1z}-k_zd}$ and $k_{1z}=-k_z$. The latter
holds because 
 $\kb$ points backward in a LHM. 
Thus with respect to the radiative decay, the second atom located in 
region 2 at $\rb^\prime$, i.e. 
on the other side of the Veselago-Pendry lens,
 behaves as if it would be located
in region 0 at position $\rb^\prime+2d{\mathbf{\hat z}}$. 
This implies that for an atom pair in the 
focal points
\be
\Gamma_{12}=\Gamma_{11}.
\ee
Thus the 
antisymmetric, single excited state $|{\rm a}\rangle$
has vanishing radiative decay, while the symmetric state $|{\rm s}\rangle$
decays with twice the free-space rate. I.e. there is
perfect sub- and super-radiance between the two atoms.

\begin{figure}[ht] 
  \begin{center} 
    \epsfig{file=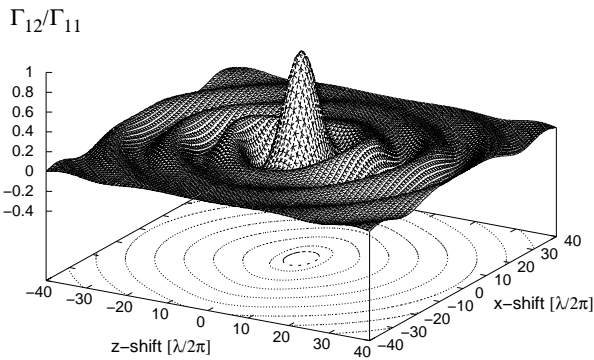,width=8cm} 
    \caption{Deviation from perfect sub/super-radiance as a function
of spatial shift of atom 2 out of focal point of atom 1. 
$z$ corresponds to axial, $x$ to radial shift. The dipole moments of the
atoms are assumed to be oriented along the {\bf x} axis.
$\Gamma_{12}/\Gamma_{11}=1$
corresponds to perfect sub-radiance of antisymmetric state, 
$\Gamma_{12}/\Gamma_{11}=0$ to independent atoms.}
    \label{fig2} 
  \end{center} 
\end{figure}

Remarkably the existence of sub-/super-radiant states does not seem to depend 
on the distance between the atoms. In particular in contrast to
free space 
% in three or one spatial dimension
% (corresponding to a atoms in a one-dimensional waveguide)
the phenomenon is possible also over distances large compared to the 
transition wavelength. This is due to the vanishing optical length of all
pathways between the two foci of an ideal Veselago-Pendry lens
 within the relevant
spectral width. 
Fig.\ref{fig2} shows the ratio $\Gamma_{12}/\Gamma_{11}$ as a function of the
spatial shift of atom 2 from the image of atom 1 at 
$x_{\rm im}=0, z_{\rm im}=0$ 
in radial ($x$) and axial direction ($z$). The profile is identical 
to the free-space case with atom 1 being located at $x=0, z=0$.
One recognizes from fig.\ref{fig2} 
that like in free space the positions of the two atoms have to be controlled
to within a fraction of the transition wavelength $\lambda$.

%%%%%%%%%%%%%%%%%%%%%%%%%%%%%%%%%%%%%%%%%%%%%%%%%%%%%%%%%%%%%%%%%%%%%%%%%%

%%%%%%%%%%%%%%%%%%%%%%%%%%%%%%%%%%%%%%%%%%%%%%%%%%%%%%%%%%%%%%%%%%%%%%%%%%

\begin{figure}[ht] 
  \begin{center} 
    \includegraphics[width=8.4cm]{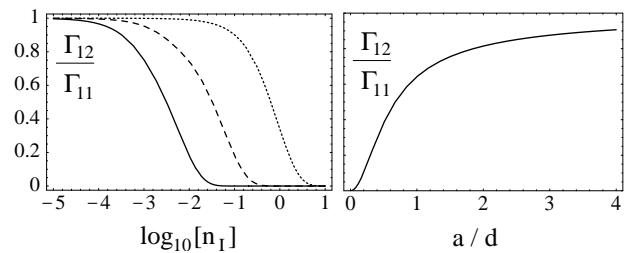}
    \caption{{\it left:} $\Gamma_{12}/\Gamma_{11}$ as function of imaginary
part of refractive index $n_I$ for Re$[n]=-1$ for different thicknesses $d$
of the lens, $d = 100 \lambda/2\pi$ (solid line), 
$d= 10 \lambda/2\pi$ (dashed), and
$d=1 \lambda/2\pi$ (dotted). {\it right:}  $\Gamma_{12}/\Gamma_{11}$
as function of transversal radius $a$.}
    \label{fig3} 
  \end{center} 
\end{figure}

When the lens is not perfect, e.g. in the presence of losses, the ratio
$\Gamma_{12}/\Gamma_{11}$ decreases roughly exponentially 
with increasing distance of the atoms
and the sub-/super-radiance effect disappears. 
This is 
illustrated in Fig. \ref{fig3}a. The radiative coupling is also
not perfect if the lens has only a limited transversal extension. It is not
possible to give an analytical expression for the Greens-tensor of 
a lens consisting of a disk of finite radius $a$. Also a numerical 
calculation of $\Gb$ for this case
is quite difficult. One can however obtain an estimate of the effect if 
$d\gg \lambda$ by employing a short-wavelength or ray-optics 
approximation. Noting that for a lossless LHM with $n(\omega_0)=-1$
only propagating waves with $k_\perp \le \omega_0/c$ contribute 
to Im$\bigl[\Gb^{20}\bigr]$, we can model the effect of a finite transverse
extension of the lens by restricting the $k_\perp$ integration 
in eq.\eqref{Green20} to
values 
\be
k_\perp \le k \frac{\frac{a}{d}}{\sqrt{\frac{1}{4} 
+\left(\frac{a}{d}\right)^2}}.
\ee
The corresponding 
result is shown in Fig. \ref{fig3}b. It is apparent that already a 
moderate ratio $a/d$
is sufficient to obtain close to 100\% suppression of decay
of the anti-symmetric state $|a\rangle$.

%%%%%%%%%%%%%%%%%%%%%%%%%%%%%%%%%%%%%%%%%%%%%%%%%%%%%%%%%%%%%%%%%%%%%%%%%%

%%%%%%%%%%%%%%%%%%%%%%%%%%%%%%%%%%%%%%%%%%%%%%%%%%%%%%%%%%%%%%%%%%%%%%%%%%

\begin{figure}[ht] 
  \begin{center} 
    \includegraphics[width=7.5 cm]{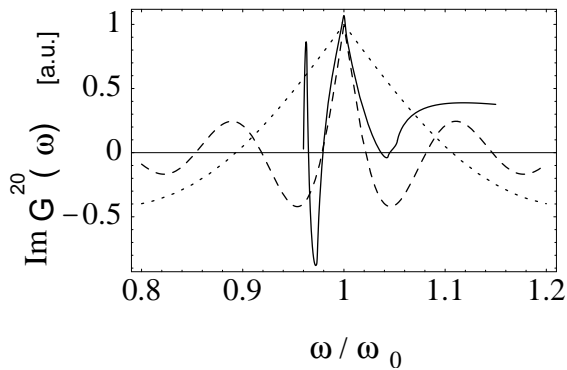}
    \caption{Im$[\Gb^{20}(\omega)]$ 
following from eq.\eqref{Green-disp}
for lossless LHM with $n=-1 +\alpha(\omega-\omega_0)$ for 
$\alpha=45/\omega_0$ for
$dk_0=1$ (dashed), $0.2$ (dotted).
Also shown is the numerically calculated spectrum for a specific 
causal model for $n(\omega)$ with resonances of
$\varepsilon(\omega)$ and $\mu(\omega)$ below $\omega_0$. 
$n(\omega)$ was chosen such that 
 Re$[n(\omega_0)]=-1$ and $\alpha=45/\omega_0$.
The central structure is well represented by the
linear-dispersion approximation \eqref{Green-disp}.
Furthermore a narrowing of the spectral width with increasing thickness
is apparent.}
    \label{fig4} 
  \end{center} 
\end{figure}

If the LHM has arbitrarily small losses at the frequency of interest and if
the lens has a sufficiently large transversal extension, the previous
discussion suggests that sub- and super-radiance is possible for
two atoms at arbitrary distance. For causality reasons this is of course
not possible. Thus the question arises what is the maximum possible separation 
$2d$ of the atoms over which the effect exists. 
As pointed out already in
the original paper by Veselago \cite{Veselago68}, 
a lossless negative index material 
is necessarily dispersive. 
The positivity of the electromagnetic energy
in a lossless LHM requires that
$\frac{{\rm d}}{{\rm d}\omega} \Bigl(\omega {\rm Re}
\left[\epsilon(\omega)\right]\Bigr) \ge 0$, and 
$\frac{{\rm d}}{{\rm d}\omega} 
\Bigl(\omega {\rm Re}\left[\mu(\omega)\right]\Bigr)\ge 0$.
which implies for $n(\omega_0)=-1:$ 
\be
\frac{{\rm d}}{{\rm d}\omega}n(\omega_0)\ge \frac{1}{\omega_0}.
\ee
As a consequence of the dispersion of the refractive index, 
the frequency window $\Delta\omega$ 
over which $\Gb^{20}(\omega)\approx \Gb^{00}(\omega)$ 
narrows with increasing thickness of the lens. When $\Delta\omega$
becomes comparable to the natural linewidth  of the atomic 
transitions $\Gamma_{11}$, the Markov approximation used 
in eq.\eqref{Liouville} is no
longer valid. To give an estimate when this happens, 
we note from eqs.\eqref{Green20}-\eqref{RS} 
that for $d\gg\lambda$ the term in $\Gb^{20}$
 that is most sensitive to 
dispersion is the exponential factor 
${\rm e}^{i\Kb\cdot(\rb-\rb^\prime)}{\rm e}^{i(k_{1z}-k_z)d}$. 
Taking into account a linear dispersion in this exponential factor, 
according to $n=-1 +\alpha (\omega-\omega_0)$, with a real value of $\alpha$, 
while keeping  the resonance values for $T^{\rm  TE}, T^{\rm TM}$
and  $R^{\rm TE}, R^{\rm TM}$, one finds 
for the Greens-tensor
\be
{\rm Im}[\Gb^{20}(\omega)]=\frac{k}{8\pi}
{\rm Re}\left[\int_0^1 \!\!{\rm d}\xi\, (1+\xi^2) {\rm e}^{i 
\frac{d k_0}{\xi}\alpha (\omega-\omega_0)}\right]\mathbf{\hat 1}.
\label{Green-disp}
\ee
As can be seen from Fig. \ref{fig4} the spectral width $\Delta\omega$ of 
the Greens-function is in this approximation
of order $\Delta\omega \approx (k_0 d \alpha)^{-1}$.
Since as mentioned above for a lossless LHM $\alpha \ge 1/\omega_0$, one 
arrives at $\Delta\omega\le {c}/{d}$.
This leads to an upper bound
for the distance of the atoms. The requirement $\Delta\omega\gg \Gamma_{11}$
leads to
\be 
d \ll \frac{c}{\Gamma_{11}}.
\ee
This condition can easily be understood. It states that the distance
between the two atoms must be small enough such that the travel time
of a photon from one atom to the other is small compared to the 
free-space radiative lifetime.

%%%%%%%%%%%%%%%%%%%%%%%%%%%%%%%%%%%%%%%%%%%%%%%%%%%%%%%%%%%%%%%%%%%%%%%%%%

%%%%%%%%%%%%%%%%%%%%%%%%%%%%%%%%%%%%%%%%%%%%%%%%%%%%%%%%%%%%%%%%%%%%%%%%%%

The existence of the sub-radiant state $|a\rangle = 
(|12\rangle-|21\rangle)/\sqrt{2}$
can be used e.g. to prepare a maximally entangled state between the two atoms
in a similar way as suggested in \cite{Beige-Knight-PRL} for a cavity system.
Preparing the first atom in the excited state, i.e. $|\psi_0\rangle
=|21\rangle = (|s\rangle +|a\rangle)/\sqrt{2}$ and letting the system evolve
leads to a 50/50 mixture of both atoms being deexcited and both atoms being in
the anti-symmetric state
$|a\rangle$, which is maximally entangled.
\be
|\psi_0\rangle\langle \psi_0|\, \longrightarrow\, 
\frac{1}{2} |11\rangle\langle 11| +\frac{1}{2}|a\rangle\langle a|.
\ee
By detecting spontaneously emitted photons it is then possible
to postselect with 50\% success probability a maximally entangled pair of 
atoms.

While the decay properties are determined 
only by the Greens-tensor at one frequency, 
the dipole-dipole
shift $\delta\omega$ depends on the whole spectrum of the dielectric 
function $\varepsilon(\omega)$ and the magnetic permeability $\mu(\omega)$.
Using various single-resonance model functions for $\varepsilon$ and
$\mu$, which fulfill Kramers-Kronig relations we found values of
$\delta\omega$ of up to $0.5 \Gamma_{11}$. It is not clear however
what the upper value of $|\delta\omega|/\Gamma_{11}$ is, as this
requires an optimization over all possible (i.e. causal) material
responses $\varepsilon(\omega)$ and $\mu(\omega)$. If
$|\delta\omega|\gg\Gamma_{11}$ could be achieved, an almost perfect
coherent excitation transfer between two atoms in the focal points of the
Veselago-Pendry lens is possible without the use of a resonator. In the opposite case
of $|\delta\omega|\ll \Gamma_{11}$ the fidelity of such a transfer 
process drops to 25\%. A more detailed study of the dipole-dipole shift
in LHM will be the subject of further studies.

In summary we have shown that two atoms put in the focal points of an ideal,
i.e. lossless Veselago-Pendry lens exhibit perfect sub- and super-radiance
as long as their distance is much smaller than the propagation length of light
corresponding to the free-space decay time. Since the latter can be
orders of magnitude larger than the wavelength, sub- and super-radiance can
occur over distances large compared to the resonance wavelength. 

J.K. acknowledges financial support by the Deutsche Forschungsgemeinschaft
through the 
GRK 792 ``Nichtlineare Optik and Ultrakurzzeitphysik''.

\def\etal{\textit{et al.}}

\end{document}